# Privacy and data balkanization: circumventing the barriers

Bernardo A. Huberman[1] · Tad Hogg[2]

**Abstract**
The rapid growth in digital data forms the basis for a wide range of new services and research, e.g., large-scale medical studies. At the same time, increasingly restrictive privacy concerns and laws are leading to significant overhead in arranging for sharing or combining different data sets to obtain these benefits. For new applications, where the benefit of combined data is not yet clear, this overhead can inhibit organizations from even trying to determine whether they can mutually benefit from sharing their data. In this paper, we discuss techniques to overcome this difficulty by employing private information transfer to determine whether there is a benefit from sharing data and whether there is room to negotiate acceptable prices. These techniques involve cryptographic protocols. While currently considered secure, these protocols are potentially vulnerable to the development of quantum technology, particularly for ensuring privacy over significant periods of time into the future. To mitigate this concern, we describe how developments in practical quantum technology can improve the security of these protocols.

**Keywords** Data privacy · Negotiation · Quantum security

## 1 The challenge of protected data

Digital data collected, stored and processed by many organizations throughout the world are often key assets for their businesses. This leads them to protect their data as a major competitive advantage. In addition, countries or regional groups of countries such as the EU are increasingly mandating restrictions on how data can be shared by these organizations, particularly with those in other jurisdictions. These restrictions arise from broad concerns that misuse of this data poses both to individuals and to nations or societies at large [1].

However, there is also a clear value in having much of that data widely shared for purposes that benefit all, such as medical research, the discovery of demographic trends and technological innovation, to name a few [2].

And yet, present trends in both national legislation and corporate attitudes are tilting the balance to more stringent privacy rules, which not only affect institutions interested in accessing pluralistic databases but also address the natural desire of corporate needs to keep data private from competitors.

A different and equally significant challenge is posed by online interactions and the massive amounts of information collected by institutions and some individuals. The challenge is manifested when a group of institutions or countries wish to work together to benefit from synergies among their different data sets. Each country has large amounts of data about its people, their demographics, medical history, prescribed treatment and outcomes, entertainment preferences, educational backgrounds, and technical data. This data contains a wealth of information that if shared or purchased by some members of the group could mutually benefit all parties. The question then arises as to how to exchange this data in such a way so as satisfy the privacy constraints imposed by different countries and institutions [3].

For example, in the case of medical data, shared information could enable faster diagnosis and more effective treatment for similar cases. Equally important, there is an opportunity for massive scale "virtual clinical trials" by combining data from different groups, with the caveat that protocols are similar enough to allow merging outcome data. This is a case where groups may need detailed information to get the full benefits of the data rather than just broad statistical summaries. However, this data usually contains extremely sensitive and private information both about patients and the

✉ Tad Hogg
tadhogg@yahoo.com

Bernardo A. Huberman
b.huberman@cablelabs.com

[1] CableLabs, Louisville, USA

[2] Institute for Molecular Manufacturing, Palo Alto, CA, USA

hospitals. Thus, for a variety of reasons—including regulatory ones—sharing this data can be problematic.

One approach to addressing privacy concerns is de-identifying the data, i.e., redacting parts of the data that make it personally identifiable. A policy choice to exploit this possibility is to not treat organizations with de-identified data as having personal data if they lack the key to re-identify the data [4]. This can be effective when the aspects of data the organizations require do not themselves allow reconstructing the identifications with high likelihood.

A different constraint on sharing in medical contexts arises due to anti-trust legislation. To prevent market collusion, the law prohibits competitors from sharing non-public information about their costs, price structure and production methods. In addition, sharing practices treated as trade secrets, rather than protected by patents, could harm the sharing company by giving up competitive advantages. However, sharing production knowledge can have a public benefit of rapidly increasing the capacity of an industry as companies learn best practices from each other. This is particularly relevant when trying to rapidly expand manufacturing capacity, such as in producing vaccines for a pandemic, where no single company has sufficient capacity to meet the market demand [5].

Ad hoc exceptions to anti-trust law can be, and have been, granted in such cases. But that only addresses the problem in high profile cases and after the need becomes widely recognized. This can delay or prevent obtaining such exceptions in smaller, more specific medical (or other) situations where the benefit only affects a small number of people, or as soon as some participants become aware of the issue. Private data matching provides an alternative that could be allowed in advance as a matter of public policy: allowing companies to use private data comparison to check for potential benefits of sharing more extensive information, then limiting the sharing to that information alone. Or, if the actual sharing still requires a case-by-case evaluation, the initial private comparison could indicate to the companies whether there is enough benefit from sharing to make a case for an exception to the rules.

The above scenarios envision multiple parties, each having a portion of the data and trying to decide whether there is a mutually beneficial opportunity to combine the data. Another scenario is one organization having data for sale that people usually access a few items at a time. This can make it prohibitively expensive to evaluate aggregate properties of the data and not just a few individual items. An example involves public judicial records: aggregate information is required to identify inconsistencies, biases, etc. in the judicial process [6]. Interested people (e.g., academic researchers) could join together to query a data set to see if the aggregate information is of interest to the group, but without getting the data itself. If they find there's something of interest, they could then obtain funding, e.g., via social research grants, to pay for the data. Otherwise, they know not to bother.

The ability to determine whether data is of interest, prior to purchase, will help more groups identify their interest in the data and thereby increase their willingness to purchase a large portion of the data.

Due to the specialized interests of potential consumers of the data, a simple 'one size fits all' summary of the data, or a few samples will not be sufficient to determine how suitable data is suitable for these interests, leading potential bulk data customers to forego the opportunity or bid much less than the data might actually be worth to them.

On a smaller scale, this problem arises with researchers deciding whether to purchase technical articles behind a paywall. If you have a specific question, viewing just an article's abstract prior to purchase may not be sufficient to decide whether the article answers your question. So instead of paying, people skip that article and look for others readily available, even if they may not be as relevant.

To support more complex and specialized evaluation of the data requires a more involved protocol, as we describe here. The data holders could be motivated to enable this protocol by the possibilities of occasional much larger purchases than their normal sales of individual cases.

To summarize, this general class of problems arises when a dataset containing private information consists of parts that belong to multiple parties or owners and they collectively want to perform analytic studies on the entire dataset while respecting the privacy and security concerns of each individual party. This is broadly referred to as privacy-preserving data mining (PPDM) or secure multi-party computation (SMC) in the literature.

## 2 Privacy-preserving mechanisms

A common approach to enable secure record linkage is to use a trusted third party (honest broker) [7, 8] or a semi-trusted third party [9, 10]. However, such solutions are often not secure [11, 12] and it may be difficult for all parties to agree on a trusted intermediary, especially if they are constrained by legal requirements, e.g., that different parts of the data must remain in different jurisdictions. To address this issue, several solutions have been proposed in the literature. Some of these solutions are based on secure protocols such as garbled circuits [13] and oblivious transfer [14]. Though these solutions provide strong security guarantees, they are inherently complex and are often restricted to a two-party scenario. On the other hand, hash-based approaches such as Bloom filters that have been proposed as alternative scalable solutions for privacy-preserving record linkage [9, 15] are susceptible to different types of attacks such as a dictionary or frequency-based ones [16, 17].

Recently there have been more direct and successful approaches. Bellala and Huberman [18] proposed a secure solution for data mapping and data linkage, which arises as a pre-processing step in a multi-party distributed data analytics task. The goal is to identify the correspondence between entities in a distributed dataset and to do so while respecting the privacy of the data.

For instance, in the healthcare domain, each hospital may have data belonging to a subset of patients with a subset of attributes.

In any multi-party distributed analytics application, one of the first steps is to ensure that the datasets and the corresponding data elements are aligned to facilitate subsequent analytics tasks such as similarity search, clustering, outlier detection, etc.

For instance, say Party 1 may want to find patients similar to Patient ID 002 in Party 2′s database. Party 1 would need to compute the similarity between this patient with all patients in the database of Party 2. To compute this similarity, Party 1 would first need to identify the set of common attributes between the two databases, and order (or link) these common attributes to facilitate similarity computation.

One approach is based on a ring protocol that works as follows.

1. Party 1 would first mask its private list $X_1$ composed of elements $\{a,b\}$ using its secret key $k1$, share it with party 2, who in turn would further encrypt the incoming data using his secret key $k2$ and share with the next party who repeats this process. The ring protocol would terminate at party 1, after all the parties have masked the data using their secret key.
2. Party 1 would publish its encrypted data to all other parties participating in the protocol.

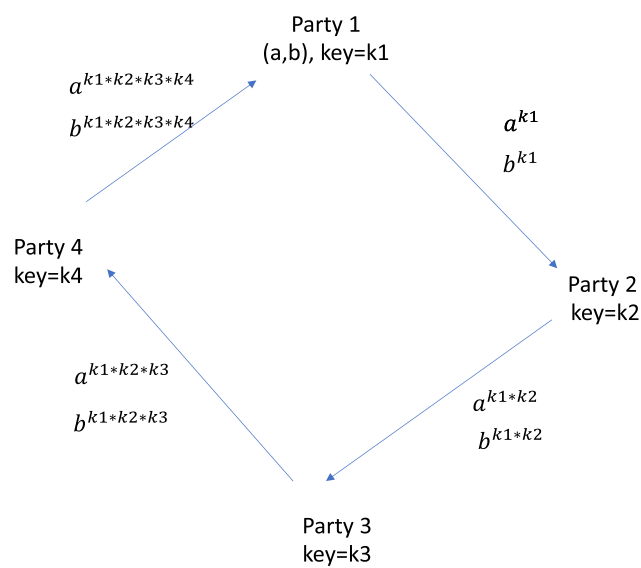

Each party would follow the same approach by applying the ring protocol to encrypt their private list using the secret keys of all parties, and then share the encrypted data with all parties. Once all parties complete the above two steps, they can find the intersecting set by matching the encrypted data and agree on a common order.

Note that this ring topology approach is not susceptible to collusion. For example, if parties $i$-1 and $i+1$ collude, they cannot guess the secret key of party i, due to the intractability of the discrete logarithm problem as described above.

An alternative approach to the second step described above would be to use an untrusted mediator (or a broker), where each party would send its final encrypted values to the untrusted mediator, who would extract the set of common entities, and order them. Note that the mediator only has access to encrypted data. Moreover, the mediator would not be able to guess the secret key of a party, even if he colluded with one or more parties, again due to the intractability of the discrete logarithm problem.

## 3  Is there room to negotiate?

Arranging to combine data while ensuring privacy requires considerable effort to set up, e.g., to gain regulatory approvals in multiple legal jurisdictions. Before taking on this effort, the parties involved would benefit from knowing whether there is room to negotiate a mutually beneficial agreement. This is a preliminary round of private data sharing where each party has their reservation price but is reluctant to reveal that price to others since that could harm their negotiating position. This reluctance can lead to extensive preliminary discussions, when there may not be an overlap of prices that could lead to a deal. Or the parties may forego attempting to find a deal because of this uncertainty.

As an example of the kinds of negotiations that can take place while keeping most of the data private is the following zero-knowledge based protocol.

A key feature of this protocol is that if the bid is not above the reservation price, no value is revealed to the seller. So here in pictorial fashion is how the method works.

# Eliminating Pointless Haggling

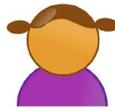 Alice wants to buy an item that Bob is selling, but does not want to reveal her bid in advance.

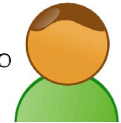 Bob is selling the item but does not want to reveal his reservation price.

Alice and Bob would like to know if they can pursue the transaction securely and privately.
Which means a) having a secured link and b) discovering if Alice's bid is higher than Bob's reservation price without revealing it.

To proceed the two parties, Alice and Bob establish a secure connection using conventional cryptographic protocols.

Once the secured communication is established, Alice and Bob encode their bids and reservation numbers respectively, in a vector such that Bob's vector components have as entries increments of its reservation number and Alice's bid vector A has one component equal to her bid and the rest of the components are 0. Alice generates a secret key x, and Bob generates another secret key y. Both Alice and Bob agree on a common large prime number p. Pictorially this is shown in this figure.

## Transaction Discovery Protocol

Alice's bid is a n-vector $A$: $A=(\$a,\$0,\$0,\ldots\$0)$
Alice generates secret key $x$

Bob's reservation is an n-vector $B=(\$r, \$5r, \$10r, \$15r,\ldots \$nr)$
Bob generates secret key $y$

- $a$'s, $r$'s, $x$, $y$ are integers.
- Alice and Bob agree on a common prime number $p$.
- All computations are done modulo $p$.

Once these steps have been taken, the following data exchanges takes place:

## Negotiation Protocol

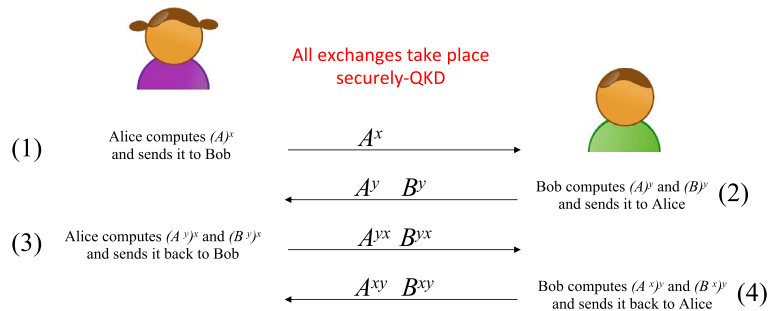

All exchanges take place securely-QKD

(1) Alice computes $(A)^x$ and sends it to Bob  →  $A^x$

(2) ← $A^y$  $B^y$  Bob computes $(A)^y$ and $(B)^y$ and sends it to Alice

(3) Alice computes $(A^y)^x$ and $(B^y)^x$ and sends it back to Bob  →  $A^{yx}$  $B^{yx}$

(4) ← $A^{xy}$  $B^{xy}$  Bob computes $(A^x)^y$ and $(B^x)^y$ and sends it back to Alice

Since $A^{xy} = A^{yx}$ they now know that Alice's bid $\$a$, matches one of Bob's list, and thus is larger or equal than Bob's reservation price.
But Alice doesn't know Bob's reservation value.

Given the above results, Alice and Bob can now transact. Notice that if Alice's bid was smaller than Bob's reservation price there would have been no match and no negotiation would be taking place. In that case, Alice walks away without knowing Bob's reservation price and Bob does not discover Alice's bid. If the transaction is feasible, Alice still does not discover Bob's reservation price and Bob knows her bid since the transaction went ahead.

The full security of the transaction is based on methods of secure multiparty computation [19].

## 4 Generalizations and limitations

Privacy-preserving mechanisms can be generalized to multiple parties. For example, to compute the average of private data values one can start with a random number, each person, in turn, adding their value to the previous one and passing the accumulated sum to the next; the last person in the group returns the accumulated sum to the first person, who subtracts the random number and divides by the number of participants. This is simpler than the cryptographic method used in the millionaire's problem.

These privacy-preserving protocols make various assumptions on the motivations of participants, with corresponding levels of complexity to prevent misuse. The simplest case is the 'honest but curious' user, who follows the protocol as agreed but will collect or infer data revealed by that protocol. At the other extreme are malicious actors who seek to subvert the protocol or pretend to perform the specified operations but instead do something else. Such cases require additional rounds of test and verification, leading to more computationally expensive protocols.

The level of security required depends on the scenario, including the usefulness for legal recourse after the fact if someone is later found to have violated their contractually agreed behavior. Our discussion focuses on the case where all parties have the same overall goal of finding out whether there is potential for a data-sharing deal, and sufficiently value their reputation (e.g., to participate in subsequent data sharing negotiations) so that the more complex cryptographic protocols are not necessary.

## 5 Quantum protocols

The above solutions suffer in principle from a fundamental lack of security in the form of third parties eavesdropping on the exchanges and possibly impersonating one negotiator. While standard cryptographic algorithms can protect the data in transit, very recent advances in quantum computing threaten the security of those protocols.

Thus the need for Improved security through use of quantum internet protocols which distribute keys though quantum channels that are provable—and not just algorithmically-secure. In particular, we describe the use of entangled photons and quantum key distribution (QKD).

This quantum protocol [20] replaces the step described above where the two parties establish a secure internet connection. With the quantum enhancement, they create a provably secure internet connection over the transport layer security (TLS) using quantum key distribution, as shown in the figure illustrating how this mechanism is implemented.

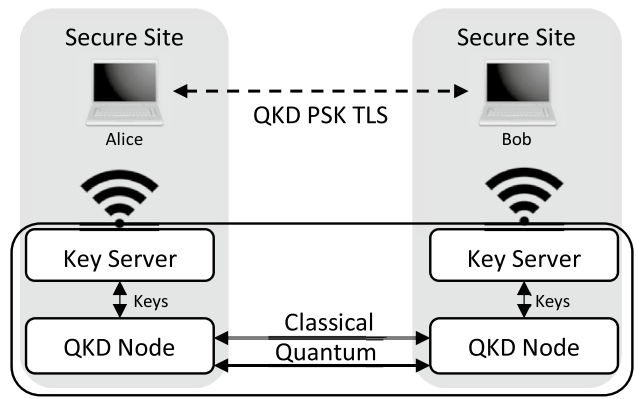

Once they have this connection, the protocol proceeds as described above.

This use of quantum technology illustrates how it could be applied to the infrastructure supporting higher-level protocols. From the user perspective, there is no change in the high-level protocol they use. This allows them to easily gain the improved security of quantum key distribution.

## 6 The promise of quantum technology

We have discussed using quantum key distribution as part of the overall protocol for secure database evaluation. As this technology improves, there will be an opportunity for quantum technology to handle a larger part of the protocol. This could further enhance security by relying on physical rather than computationally difficult problems for security. This could be particularly relevant when there are concerns that any data exchange protocol is not only secure now but also for many years into the future, e.g., because the data may need to remain secure for the lifetime of the participants in the database. Enhanced quantum technology threatens the security of common protocols in use today, so switching to alternatives could help alleviate these concerns of future compromise, thereby enabling more extensive use of protocols for identifying beneficial data sharing.

In addition to improving security, quantum technology offers the possibility of altering the incentive structure of the transactions. This application of quantum technology is distinct from its use to improve security or speed up some computations. Specifically, quantum transactions provide new possibilities on how information is exchanged and deleted, and how choices are correlated. This leads to changes in the incentive structure of games [21], which can be applied to help enable cooperative decision-making in groups [22] and improve bidding incentives via information hiding in auctions [23]. The limitations on the length of time quantum information can be stored provides alternate methods for the private data access [24].

One example is enhanced incentives for cooperation in two-party negotiations and public goods provisioning. Another example is quantum auctions that do not reveal losing bids without the need for a trusted escrow agent. This can improve incentives for truthful bidding, especially in multiple round settings, e.g., when companies expect to bid against the same competitors each year for a similar auction, such as frequency spectrum.

This could apply to data sharing that the parties expect to perform on a continuing basis, e,g., to handle updates to the data as each party continually receives new information.

As described above, removing personally identifiable information from data can help address privacy concerns in sharing the data. However, correlations among aspects of the data and specific individuals may allow re-identifying individuals, especially through the application of large-scale data analysis methods [25]. Thus even if redacted data sets are individually private, their combination may not be. In that case, protocols using homomorphic cryptography or the information hiding properties of quantum states may be helpful approaches to allowing the use of the combined data while reducing the possibilities for re-identifying people from the combination. Designing and evaluating such protocols are good directions for future investigation.

## 7 Conclusion

We described how privacy-preserving mechanisms can aid the discovery of shared interest in aggregating multiple data sets owned or controlled by different organizations. One application is using privacy-preserving methods to test for overlap between the bid and ask prices in a negotiation. These techniques could help unlock the synergies among large data sets held by different organizations in different legal jurisdictions. This could realize the potential of the significant value to be gained by combining the data. Moreover, these protocols could reduce the many concerns organizations and governments have in specifying requirements for sharing data, including legal constraints and the fear of revealing cost preferences to competitors. We described how these protocols could be enhanced by the ongoing development of quantum technology that may undermine the security of conventional methods.